\documentclass[10pt,conference]{IEEEtran}
\IEEEoverridecommandlockouts
\usepackage{cite}
\usepackage{amsmath,amssymb,amsfonts}
\usepackage{graphicx}
\usepackage{textcomp}
\usepackage{xcolor}
\usepackage{hyperref}
\def\BibTeX{{\rm B\kern-.05em{\sc i\kern-.025em b}\kern-.08em
    T\kern-.1667em\lower.7ex\hbox{E}\kern-.125emX}}

\usepackage{listings}
\usepackage{balance}
\usepackage{multicol}
\usepackage{adjustbox}
\usepackage{listings}
\usepackage{xcolor}
\usepackage{pbalance}
\usepackage{pgfplots}
\usepackage{algorithm}
\usepackage{algpseudocode}
\pgfplotsset{compat=1.12}
\usepackage{subcaption}
\usepackage{graphicx}
\usepackage{booktabs}
\usepackage{tikz}
\usetikzlibrary{backgrounds}
\usetikzlibrary{arrows,shapes}
\usetikzlibrary{arrows.meta}
\usetikzlibrary{tikzmark}
\usetikzlibrary{calc}
\usepackage[most]{tcolorbox}

\usetikzlibrary{shapes.arrows}

\tikzset{
    cross/.pic = {
    \draw[rotate = 45] (-#1,0) -- (#1,0);
    \draw[rotate = 45] (0,-#1) -- (0, #1);
    }
}

\usepackage{amsmath}
\usepackage{amsthm}
\usepackage{mathtools, nccmath}
\usepackage{wrapfig}
\usepackage{comment}
\usetikzlibrary{calc,shapes.geometric,arrows,positioning,intersections,automata, backgrounds,}

\def\code#1{\textcolor{red}{\texttt{#1}}}

\definecolor{codegreen}{rgb}{0,0.6,0}
\definecolor{codegray}{rgb}{0.5,0.5,0.5}
\definecolor{codepurple}{rgb}{0.58,0,0.82}
\definecolor{backcolour}{rgb}{0.95,0.95,0.92}

\definecolor{solablue}{HTML}{0068AD}
\definecolor{solaorange}{HTML}{FF5D00}
\lstdefinelanguage{JavaScript}{
	keywords={typeof, new, true, false, try, catch, function, return, null, catch, switch, var, if, in, while, do, else, case, break,let, const, throw, await},
	keywordstyle=\color{red}\bfseries,
	ndkeywords={class, export, boolean, throw, implements, import, this, from},
	ndkeywordstyle=\color{red}\bfseries,
	identifierstyle=\color{darkgray},
	sensitive=false,
	comment=[l]{//},
	morecomment=[s]{/*}{*/},
	commentstyle=\color{darkgray}\ttfamily,
	stringstyle=\color{solablue}\ttfamily,
	escapeinside={/*\#}{\#*/},	
	morestring=[b]',
	morestring=[b]",
	morestring=[b]`
}

\lstdefinestyle{mystyle}{
     backgroundcolor=\color{backcolour},   
    commentstyle=\color{codegreen},
    keywordstyle=\color{magenta},
    numberstyle=\tiny\color{codegray},
    stringstyle=\color{codepurple},
    basicstyle=\ttfamily\footnotesize,
    breakatwhitespace=false,         
    breaklines=true,                 
    captionpos=b,                    
    keepspaces=true,                 
    numbers=left,                    
    numbersep=5pt,                  
    showspaces=false,                
    showstringspaces=false,
    showtabs=false,                  
    tabsize=2
}

\AtBeginDocument{%
  \providecommand\BibTeX{{%
    Bib\TeX}}}

\newif\ifcomments
\commentstrue
\ifcomments

\commentsfalse

\ifcomments
    \newcommand{\crs}[1]{[{\color{blue}Cris: #1}]}
    \newcommand{\GP}[1]{[{\color{red}GP: #1}]}
    \newcommand{\gian}[1]{[{\color{violet}Gian: #1}]}
\else
    \newcommand{\crs}[1]{}
    \newcommand{\GP}[1]{}
    \newcommand{\gian}[1]{}
\fi

\newcommand{\toolname}{\textsc{Graphia}}

\makeatletter
\newcommand{\linebreakand}{%
  \end{@IEEEauthorhalign}
  \hfill\mbox{}\par
  \mbox{}\hfill\begin{@IEEEauthorhalign}
}
\makeatother
    
\begin{document}

\title{Call Me Maybe: Enhancing JavaScript Call Graph Construction using Graph Neural Networks}

\author{\IEEEauthorblockN{Masudul Hasan Masud Bhuiyan}
\IEEEauthorblockA{\textit{CISPA Helmholtz Center for Information Security} \\
Saarbrücken, Germany \\
masudul.bhuiyan@cispa.de}
\and
\IEEEauthorblockN{Gianluca De Stefano}
\IEEEauthorblockA{\textit{CISPA Helmholtz Center for Information Security} \\
Saarbrücken, Germany \\
gianluca.de-stefano@cispa.de}
\linebreakand
\IEEEauthorblockN{Giancarlo Pellegrino}
\IEEEauthorblockA{\textit{CISPA Helmholtz Center for Information Security} \\
Saarbrücken, Germany \\
pellegrino@cispa.de}
\and
\IEEEauthorblockN{Cristian-Alexandru Staicu}
\IEEEauthorblockA{\textit{CISPA Helmholtz Center for Information Security} \\
Saarbrücken, Germany \\
staicu@cispa.de}
}
\maketitle

\begin{abstract}
  Static analysis plays a crucial role in identifying many types of bugs, including security vulnerabilities.
  Constructing accurate call graphs, which model function invocations within programs, is a critical step in static analysis to serve as the foundation for interprocedural analyses. 
  However, due to hard-to-analyze language features, existing call graph construction algorithms for JavaScript are neither sound nor complete. 
  Prior work shows that even the most advanced solutions in this domain both generate spurious edges and miss call relations. 
  In this work, we aim to assist existing JavaScript call graph construction solutions by identifying missed call edges. 
  Our main insight is to model the call graph augmentation problem as link prediction on entire program graphs and to use a rich graph representation that includes multiple types of edges. 
  In this way, our approach, \toolname{}, can leverage the recent advances in graph neural networks on large graphs to capture non-local relations between program elements. 
  Concretely, we propose representing JavaScript programs using a combination of syntactic- and semantic-based edges.
  We show that \toolname{} can learn from imperfect ground truths: static call edges produced by existing tools or a mix containing also dynamic edges collected from unit tests of the target project, or even from other projects.
  Due to the sparse nature of call graphs, we argue that typical machine learning metrics like the ROC curve do not suffice, and instead, we evaluate the 
  success of \toolname{} by ranking candidate function definitions for each unresolved call site. 
  We present an extensive evaluation using 50 popular JavaScript libraries and 163K call edges (150K static and 13K dynamic ones). Moreover, when representing the target programs as graphs, \toolname{} uses 6.6M structural and 386K semantic edges. For more than 42\% of the statically-unresolved call sites, \toolname{} correctly predicts the right target function as the top candidate. Moreover, \toolname{} successfully predicts
true edges within the top 5 rank in 72\% of the cases, allowing both future analysts and downstream analyses to focus on a small list of promising candidates.
  These results show that learning-based approaches can indeed be used to enhance the recall of JavaScript call graph construction.
  Moreover, we believe that ours is the first work to show that GNN-based link prediction can be applied holistically to multi-file program representations to assist inter-procedural code analysis tasks.
\end{abstract}

\section{Introduction}

Call graphs depict the function invocations within programs, enabling the creation of interprocedural program analyses. 
The use of static call graphs in program analysis traces its roots back to 1970, when Barth et al.~\cite{barth1977interprocedural} and Graham et al.~\cite{graham1982gprof} pioneered this area. 
An ideal call graph should possess two properties: (1) soundness, indicating the ability to resolve all function call invocations, and (2) completeness, implying the avoidance of any false invocation edges. 
However, Rice's theorem~\cite{rice1953classes} asserts that achieving both these properties is generally impractical.
Consequently, existing analyses strive for reasonable trade-offs between soundness and precision while considering performance constraints. As an example, the state-of-the-art call graph construction tool WALA~\cite{WALAdown41:online} has a recall value of only 62\%~\cite{antal2023javascript} and a 70\% ratio of false edges~\cite{utture2022striking}, when applied to JavaScript programs.


This is mainly because in languages like JavaScript, where functions are first-class citizens, constructing a call graph is a chicken-and-egg problem. To resolve certain call sites, one needs a sophisticated interprocedural analysis, which in turn requires a call graph. 
To tackle this predicament, researchers advocate for improving the foundational pointer analysis, integral to most call graph construction algorithms~\cite{bravenboer2009strictly, mangal2015user, tan2016making}. 
Since achieving a flawless pointer analysis is generally impractical, as per Rice's theorem, analyses aim for a better scalability-precision trade-off. 
For instance, implementing a context-sensitive analysis with WALA for pointer analysis only marginally reduces the false positive rate (by 8.6\%) compared to a context-insensitive analysis~\cite{utture2022striking}, but significantly impacts the overall performance cost.


Recently, researchers have turned to machine learning to address challenges in static call graph construction. Utture et al.~\cite{utture2022striking} propose cGPruner, which reduces false edges by 23\% using random forest classifiers. AutoPruner by Thanh et al.~\cite{le2022autopruner} combines structural features from cGPruner with semantic features from CodeBERT and RoBERTa, using a neural classifier for enhanced call graph edge classification. 
While these approaches help reduce the number of false positives, we are not aware of any prior work that aims to reduce the number of false negatives, i.e., unresolved call sites. 
We note that industrial static analysis frameworks, such as WALA, already have a low false positive rate, since they are configured to produce precise but incomplete call graphs~\cite{antal2018static}, instead of generating numerous spurious edges. In our experiments, we measure that this decision results in 60\% unresolved call sites for CodeQL, a popular static analysis tool for JavaScript.

Therefore, in this work, we propose \toolname{}, an approach that uses machine learning to assist state-of-the-art JavaScript call graph construction in reducing the number of unresolved call sites. This is an inherently difficult problem due to the complexities of the language~\cite{chakraborty2022automatic}, e.g., higher-order functions, dynamic property accesses, and prototype inheritance. Thus, to support such advanced features, the analysis needs to consider the whole context around an unresolved call site. 
To this end, we propose representing the entire program under analysis using a graph neural network (GNN) and modeling the call graph augmentation problem as link prediction. 
Leveraging the comprehensive view provided by this holistic code representation, the GNN proves effective in capturing intricate relationships across the code base to handle advanced, non-local patterns like higher-order functions or dynamic property accesses. To the best of our knowledge, ours is the first work that proposes a learning process that acts on the entire program representation without involving an embedding step. This is analogous to the interprocedural paradigm in traditional static analysis, in which multiple functions are analyzed at once.

An alternative design would be to follow something more akin to intraprocedural analysis. That is, 
to use simpler machine learning algorithms combined with powerful code embeddings that incorporate rich structural information from the vicinity of the relevant program element, e.g., abstract syntax tree~\cite{Alon2018code2vecLD, Maddison2014StructuredGM, Zhang2019ANN}, or data flow graphs~\cite{Allamanis2017LearningTR, BenNun2018NeuralCC, Li2018VulDeePeckerAD, Zhao2018DeepSimDL, Zhou2019DevignEV}. However, a common drawback of these approaches is their limited contextual scope. The majority of existing embedding techniques only consider the code, abstract syntax tree~\cite{chen2018tree}, or data flow information~\cite{Zhou2019DevignEV} within the function bodies of callers and callees~\cite{sui2020flow2vec}. This narrow focus essentially hides crucial contextual details from the downstream models. This limitation is particularly noticeable in dynamic languages like JavaScript, where coding practices such as anonymous functions and higher-order functions are prevalent. Handling such code patterns involves managing complex interprocedural relations, e.g., tracking the passing of function pointers. This underscores the necessity for a different formulation of the learning task to consider the context of the entire code base.


There are multiple challenges to be solved before applying graph neural networks to entire programs. First, how to represent the code to take advantage of the existing GNN architectures. Code representations like abstract syntax trees tend to be sparse, contradicting the dense graphs on which link prediction is traditionally applied, e.g., collaboration networks among scientists~\cite{ZhangC18}. Thus, we propose a more dense program representation that utilizes both syntactic and semantic edges among the program elements. 
Concretely, we connect together nodes that refer to the same identifier.
This enhancement of the AST forces parts of the code that handle the same concept closer together, facilitating the GNN's ability to trace likely data flow relations. 

Another critical challenge is obtaining training data, considering the limitations of existing approaches and the lack of good data sets in this space. We propose extrapolating from the static edges produced by the call graph generation tool we aim to extend. Such tools are cheap to run, but as discussed above, they might miss a lot of valuable edges. GNNs were shown to be effective at link prediction in other scenarios where the ground truth is limited. For example, Zhang and Chen~\cite{ZhangC18} remove 10\% of the edges in the ground truth and use them as a test set.
Thus, we hypothesize that using solely such static edges, \toolname{} can approximate the computation of these tools and generalize the link prediction to blind spots of the tools, e.g., caused by bugs. However, the very difficult unresolved call sites are fundamentally out of the distribution points for this training setup since the baseline tool cannot produce such links. For such cases, we propose extracting dynamic edges using instrumented executions of unit tests and including such edges in the training mix. We also use dynamically extracted edges as ground truth for the experiment, extracted either from the target project or from a different one. 


We evaluate \toolname{} on the call graph of 50 popular npm libraries. For each statically unresolved call site, we judge the model's success in ranking all potential candidate edges from the call site to any other function definition in the project. Our evaluation reveals that in 42\% of cases, \toolname{} places the correct edge on the first position in the ranking, and in 72\% of instances, it predicts the correct edge within the top five ranks. To further understand the impact of our design choices, we conduct two ablation studies. The first study explores the effect of our suggested code representation, demonstrating a performance drop of 61\% without its inclusion. The second study investigates the impact of our proposed node features, revealing a 13\% performance drop when these features are omitted. Thus, the results show that \toolname{} is able to assist the existing call graph creation solutions with resolving difficult call sites. They also show that applying graph neural networks to entire programs is feasible, allowing the model to perform intraprocedural inferences.

In summary, our contributions are as follows:
\begin{itemize}
    \item We present \toolname{}, an approach for improving the recall of existing JavaScript call graph creation methods. Concretely, we model the call graph enhancement problem as link prediction with graph neural network, leveraging both structural and semantic information from code, to handle unresolved call sites.
    
    \item To the best of our knowledge, we are the first to show that link prediction using graph neural networks is feasible on holistic, multi-file program representations.
    
    \item Through extensive experiments, we show that \toolname{} can learn from imperfect ground truths, i.e., either directly using the static edges produced by a baseline static analysis tool or a combination of such static edges and dynamic ones extracted from unit test executions.
    
    \item We conduct an ablation study to gain deeper insights into our approach, affirming the effectiveness of our proposed code representation.
\end{itemize}

\section{Motivating Example}
In Figure~\ref{fig:caller} and ~\ref{fig:callee}, we present an illustrative example to highlight the rationale behind our approach and underscore the limitations of existing methods that exclusively consider the code in the vicinity of the call site. In line 1, an object of the \code{lexer} type is instantiated and assigned to the variable \code{lexer}. Advancing to line 4, we observe the invocation of the \code{showPosition} property of the lexer object. Notably, this property is not locally defined but implemented as a function elsewhere in the source code (Shown in Figure~\ref{fig:callee}). Resolving the call site via property resolution requires resolving the type of \code{this.lexer}, which lies beyond the caller function's scope. Concretely, the anonymous callee function in line 3 of Figure~\ref{fig:callee} is attached to a named property of a local object literal. This object is then returned from an immediately invoked function and assigned to a \code{lexer} property of a parser object. In line 1 of Figure~\ref{fig:caller}, such a parser object is referenced via \code{this} and instantiated via prototype inheritance. It then serves as the base object in the invocation of \code{showPosition}.
This example highlights the challenge posed by JavaScript's dynamic features. To reason about such complex call sites, an analysis needs to trace property reads and writes, reason about complex types constructed via idiomatic object initialization, and understand prototype inheritance. All these abilities are far from trivial and lie beyond the reach of most static JavaScript analyses.

The initialization of the \code{lexer} object outside the function boundary on line 1 makes it difficult to infer the complex relationship between the caller and the invoked object's property. Models like AutoPruner~\cite{le2022autopruner}, leveraging CodeBERT, may struggle to capture these nuances without considering the broader code context and runtime information. Recognizing that the \code{showPosition} property is essentially implemented as a function, requires a deeper understanding of the surrounding code and its dynamic behavior. Models lacking this contextual awareness may encounter difficulties inferring these intricacies, underscoring the significance of our proposed approach.



\begin{figure}
\centering
\begin{adjustbox}{width=\linewidth}
\begin{lstlisting}[language=JavaScript]
var lexer = Object.create(this.lexer);
if (lexer.showPosition) {
    errStr = 'Parse error on line ' + (yylineno+1) 
        + ":\n" + lexer.showPosition() + "\nExpecting "
        + expected.join(', ') + ", got '"
        + (this.terminals_[symbol] || symbol) + "'";
}
\end{lstlisting}
\end{adjustbox}

\caption{Caller function code snippet from \it{formula-parser} JavaScript library}
\label{fig:caller}
\centering
\begin{adjustbox}{width=\linewidth}
\begin{lstlisting}[language=JavaScript]
var lexer = function (){
  var lexer = ({
    showPosition: function () {
      var pre = this.pastInput();
      var c = new Array(pre.length + 1).join("-");
      return pre + this.upcomingInput() + "\n" + c + "^";
    }
  });
  return lexer;
})();
parser.lexer = lexer;
\end{lstlisting}
\end{adjustbox}
\caption{Callee function code snippet from \it{formula-parser} JavaScript library}
\label{fig:callee}
\vspace{-5mm}
\end{figure}

\begin{figure}
\begin{center}
  \includegraphics[width=\linewidth, keepaspectratio]{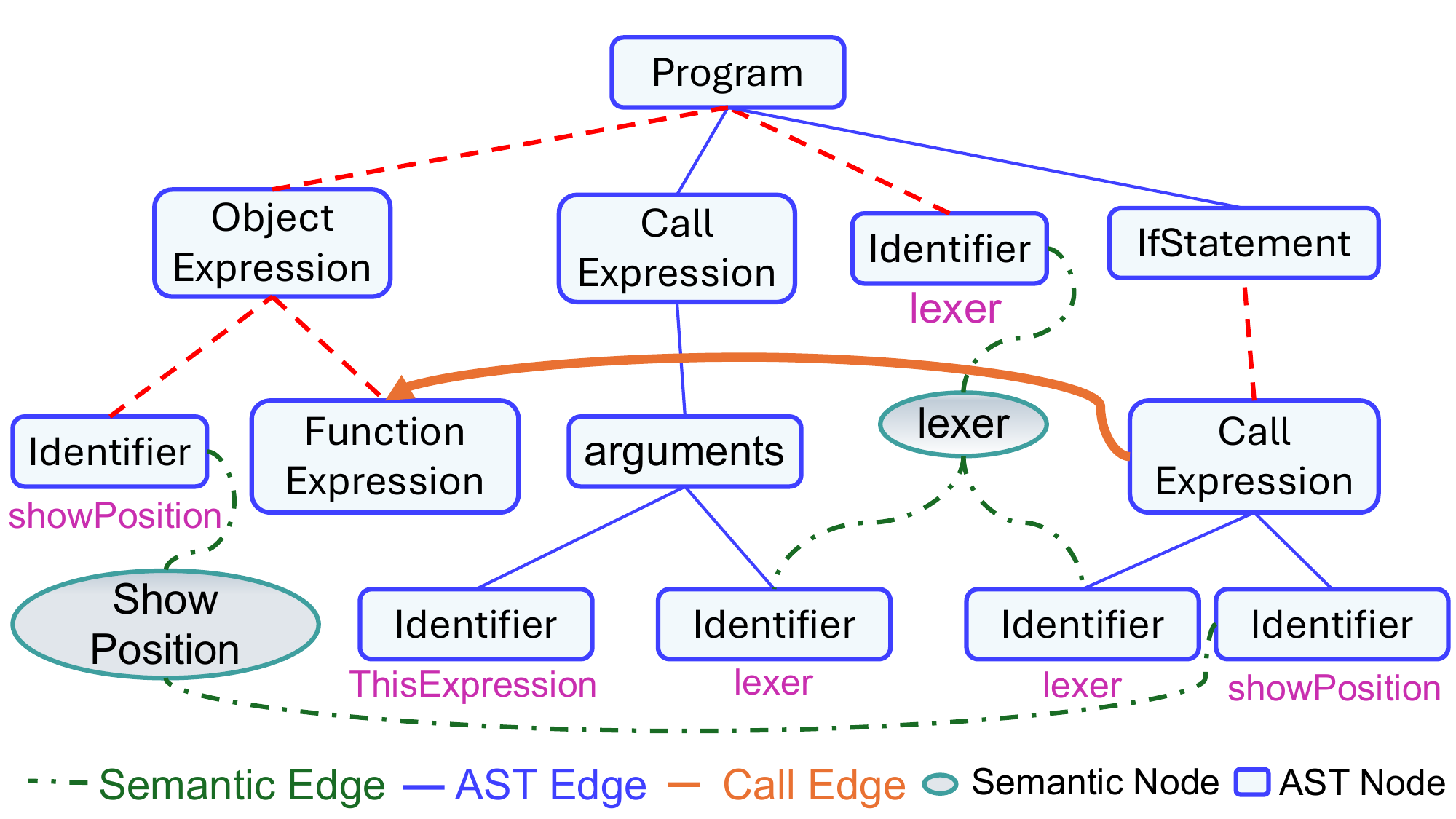}
\end{center}
\vspace{-3mm}
\caption{Combined code representation containing both syntactic and semantic nodes and edges. The graph corresponds to the code examples in Figure~\ref{fig:callee} and Figure~\ref{fig:caller}, but it was simplified for conciseness.}
\label{fig:code-repr}
\vspace{-5mm}
\end{figure}

To illustrate how our proposed approach can overcome the limitations of existing methods, let's examine Figure~\ref{fig:code-repr}. 
We represent the program's source code as a graph, incorporating different edge types to model both syntactic and semantic relationships between various nodes. The foundation of our program graph is the abstract syntax tree, produced using an off-the-shelf parser. Each syntax node is labeled with the associated node type produced by the parser. By employing structural representations and simulating data flow information through semantic edges, modern graph neural network models can adeptly infer long-distance relationships in interprocedural analysis. It is crucial to acknowledge that GNNs traditionally encounter challenges in leveraging long-range interactions, as their learned representations often tend to be locally focused. However, by incorporating semantic edges, we establish connections between related code elements, empowering the model to capture long-distance interprocedural relations effectively. This integration of semantic edges within the GNN framework fosters a deeper understanding of code dependencies and augments the ability to analyze intricate relationships across different code segments.


In the provided example, our model can extrapolate that when the \code{showPosition} method of the \code{lexer} object is invoked, it effectively calls a member method of that same object. Despite the deferred definition, these components are interconnected through semantic edges linking to \code{showPosition} and \code{lexer} nodes. The \code{showPosition} semantic node in our code representation (highlighted in green) enables the GNN model's ability to discern this non-local relationship. GNNs leverage both the code structure and feature vectors associated with each node to extrapolate this intricate relationship. This illustrates how our proposed approach enhances the model's capacity for a more nuanced and comprehensive understanding of complex relationships within the code base.

\section{Methodology}
\begin{figure}
\begin{center}
  \includegraphics[width=.9\linewidth, keepaspectratio]{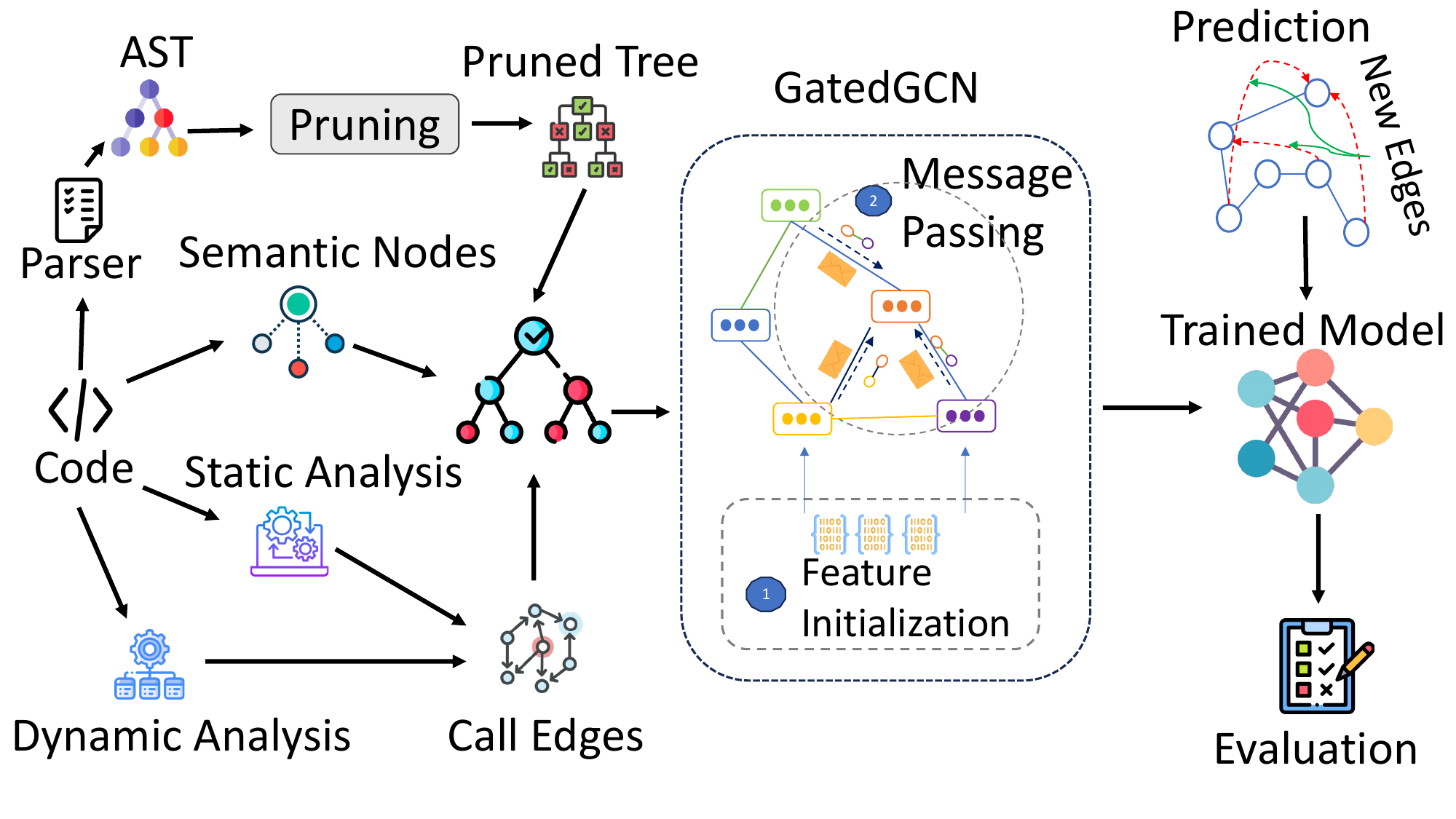}
\end{center}
\vspace{-3mm}
\caption{The overview of \toolname{} framework.}
\label{fig:overview}
\vspace{-5mm}
\end{figure}

Figure~\ref{fig:overview} illustrates the overall structure of our proposed approach. We begin by parsing the code using an off-the-shelf parser to generate an abstract syntax tree. We employ a pruning algorithm to reduce the size of this tree, selectively removing non-essential nodes while preserving the tree's overall structure. Semantic nodes are then introduced for distinct identifiers and then connected via semantic edges to each of their usages inside the syntax tree. This enables the model to reason about identifiers' usage and potentially track data flows. The pruned AST, enriched with semantic information, is now a graph, which is input to the graph neural network. 
Each node is initialized with a feature vector, and the model is trained using both static- and dynamically-derived edges. 
We use existing static call graph creation tools and instrumented execution of unit tests to collect the ground truth call edges, which we connect to the underlying code representation.
Post-training, the model predicts new edges for statically unresolved call sites. For each such call site, \toolname{} produces a list of ranked function definitions to which the call resolves. We evaluate the model's success by analyzing how high in the ranking it places hard-to-analyze call edges.

\subsection{Generating Pruned AST}

The first step in our methodology involves parsing the source code using a JavaScript parser. The parser transforms the raw code into an abstract syntax tree (AST), a hierarchical data structure that represents the syntactic structure of the program. This AST serves as the foundation for our subsequent analysis and modeling. In theory, this code representation suffices to enable a sophisticated machine learning model to perform arbitrary complex analysis tasks. In practice, however, this representation is too low-level, containing many details about the syntax of the programs instead of focusing on semantics.
Recognizing this challenge, we implement a carefully designed pruning algorithm that selectively removes intermediary non-terminal nodes, such as those representing expression statements, binary expressions, and literals. The goal is to reduce the size of the tree while preserving its overall structure and connectivity. 

The pruning algorithm operates on the AST's graph representation \( G = (V, E) \), using a parentChildDict to map node relationships. It iterates through each node \( v \), evaluating it against predefined criteria for removal. Qualified nodes are disconnected from their parents, their children are reassigned to the parent node, and the \(\text{parentChildDict}\) is updated to maintain the connectivity of the tree's structure. For instance, when an expression statement contains multiple sub-binary expressions, all child binary expressions are removed, and their children are transferred to the parent expression statement. The process reduces AST size, preserving critical node relationships and leading to 60\% faster model training.

\subsection{Increasing Connectivity via Identifiers}
Our second strategy is the introduction of semantic nodes into the pruned AST to enrich the graph's connectivity and reduce the average distance between nodes. These semantic nodes, representing distinct identifiers within the codebase, are connected to corresponding identifier usage nodes through semantic edges, depicted by green lines in our visualizations. These edges carry variable weights during training, allowing the model to discern and utilize semantic relationships more effectively. Integrating semantic nodes and edges into the pruned AST creates a comprehensive graph structure that encapsulates the source code's syntactic and semantic characteristics. We hypothesize that semantic nodes and edges allow the model to relate parts of the model that handle the same semantic elements, e.g., the same method name or global variable.

\subsection{Generating Call Edges Using Static and Dynamic Analysis}
\label{}
Obtaining reliable ground truth is vital for effective model training, especially in the domain of program analysis, where both the precision and completeness of data are crucial. Traditional tools often struggle to meet these requirements simultaneously, leading to a common situation where tools sacrifice completeness for the sake of precision~\cite{antal2018static}. To accommodate this reality, we adopt popular static analysis tools to generate accurate analysis data. Nonetheless, while precise, these tools miss a lot of edges; thus, the training data represent an under-approximation of the actual call graph. This data, while not capturing the entire graph, reliably represents true call relationships within a codebase. To generate negative examples, we choose random pairs of call sites and function definitions for which there is no edge in the ground truth.


To complement the statically-extracted data, we also perform a dynamic analysis on those programs in the data set with good unit tests. For this, we utilized the test scripts that come with these programs, instrumenting them using source-to-source transformations to capture dynamic call edges, i.e., at each function entry point, we log the call stack to identify the invoking call site. A shortcoming of using existing unit tests is that they might not execute the entire program functionality, leading to still incomplete call graphs. Despite this limitation, the combination of static analysis and the dynamic insights from test executions allows us to study how the different types of call edges (static or dynamic) influence the performance of \toolname{} and simulate realistic scenarios in which the training data is incomplete.

\begin{table}[h]
\centering
\begin{tabular}{l l} 
 \toprule
 \emph{Feature} &  \emph{Description}\\
\midrule
node\_type & Type of the AST node\\
 name & Associated identifier  for the AST node\\
 number\_of\_parameter &  Number of parameters for callee\\
 number\_of\_argument & Number of arguments for caller\\
 \bottomrule
\end{tabular}
\caption{Feature set of GNN model.}
\label{table:features}
\vspace{-5mm}
\end{table}

\subsection{Training and Evaluation}

Inspired by the feature selection criteria outlined by Utture et al.~\cite{utture2022striking}, we initialize each node in our graph neural network model with a feature vector comprising four selected features that model the surrounding program context. We show these features in Table~\ref{table:features}, and they capture the corresponding AST node type and name, together with the number of arguments or parameters for function invocation or definition nodes. These features are indispensable during the training and inference stages of the model, as discussed in Section~\ref{sec:ablation}. Our approach harmoniously combines syntactic simplification through pruning with semantic enrichment via semantic nodes and edges. This structured integration forms the foundation of our GNN input, enabling efficient information propagation and learning across nodes, thereby significantly enhancing the model's performance in program analysis.

Following the construction of the input graph and initializing nodes, the graph neural network undergoes a dedicated training phase. The model adeptly learns intricate relationships within the program graph by leveraging the approximate ground truth discussed earlier. Through iterative optimization of model parameters during training, the GNN enhances its accuracy prediction capacity. Post-training, the model is deployed to predict new edges within the program graph. 
Concretely, we train a graph neural network for each target program using a single dataset. We split this dataset into three subsets: 80\% training, 10\% validation, and 10\% testing. We use five layers for our graph neural network. In our training, we set a maximum of 500 epochs, with a stipulation to cease training if the learning rate falls below \(1 \times 10^{-5}\). Simultaneously, we maintain a batch size of 32,768 for each epoch, a figure strategically chosen to optimize both computational efficiency and training effectiveness. During each epoch, we select all edges in the validation set as true positives, and to ensure balance, we include an equivalent number of negative edges. These negative edges are composed of random pairings between call sites and function definitions that do not have an existing edge in the ground truth.


\subsection{Evaluation Metrics}

Assigning appropriate evaluation metrics is crucial for validating machine learning pipelines. While standard metrics like precision, recall, and F1-score are commonly used, they may be misleading for call graph generation due to dataset imbalance and failure to consider the entire candidate space. The ROC curve, another prevalent method, also has limitations in this context. Call graph datasets often have significantly more negative edges than positive ones, and traditional metrics don't account for the extensive candidate space in real-world scenarios. This reliance on conventional metrics may not accurately reflect the model's performance in practical applications.

\begin{figure}
\begin{center}
      \includegraphics[width=.45\linewidth, keepaspectratio]{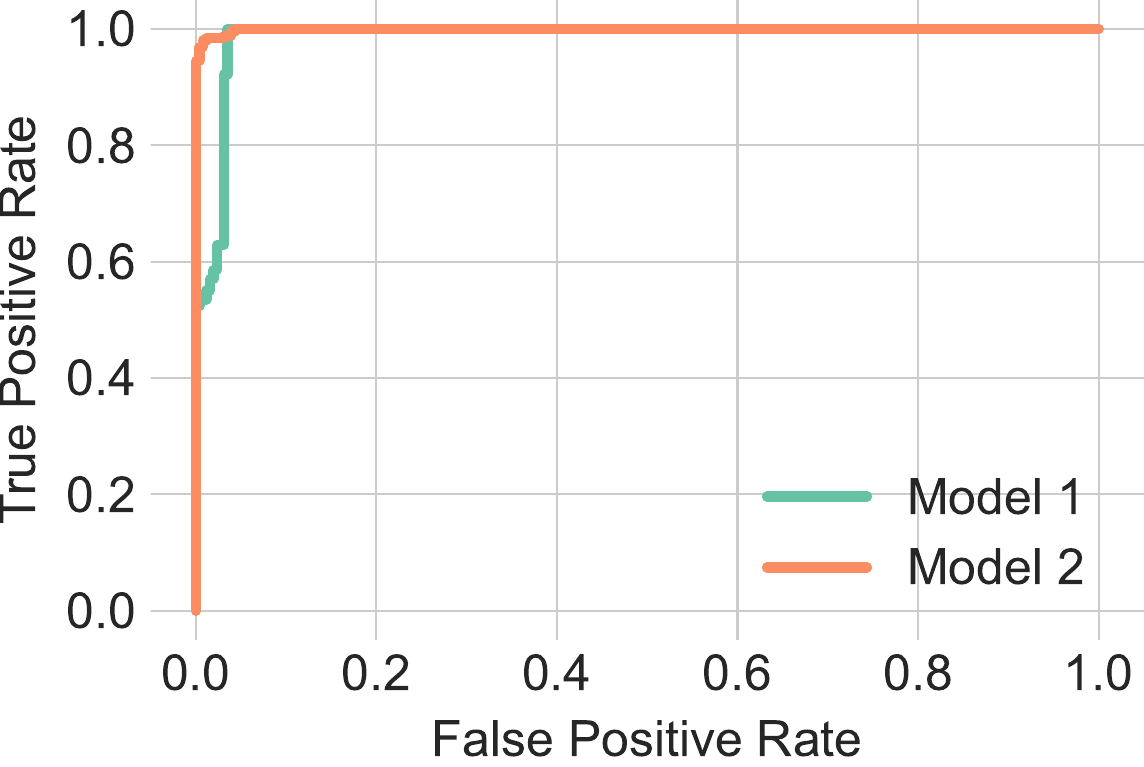}
      \includegraphics[width=.45\linewidth, keepaspectratio]{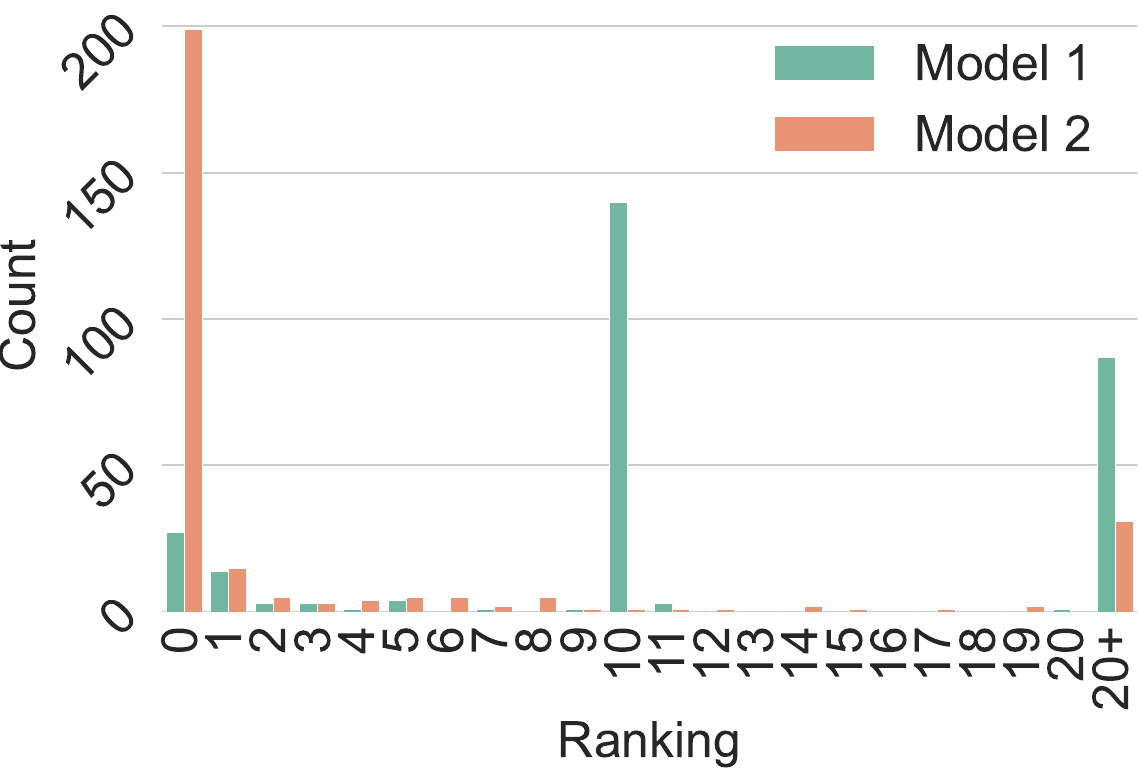}
\end{center}
\vspace{-3mm}
\caption{Performance of two models with similar ROC curves, but very different callees ranking ability.}
\label{fig:roc-ranking}
\vspace{-5mm}
\end{figure}

To illustrate this further, let's examine the ROC curve alongside Precision, Recall, and F1-score metrics for two distinct models. In Figure~\ref{fig:roc-ranking}, the ROC curve illustrates the comparative performance of two models. Despite the visual indication that Model 2 outperforms Model 1, a granular examination using precision, recall, and F1-score at a 95\% true positive rate presents a paradox. The point corresponds to 0.89 threshold value for Model 1, and 0.58 for Model 2. This means that any edge with a probability score higher than these values will be recommended as an output call edge.
Model 1 yields 96\% precision, 98\% recall, and 97\% F1-score, while Model 2 boasts 100\%, 94\%, and 96\%. However, these numbers do not reflect the underlying data distribution in our domain. 
Let us now consider the results in Figure~\ref{fig:roc-ranking}, where we depict the performance of the model configuration above on the test set. For each call site, we ask the model to produce the highest 20 candidate predictions, sort this list, and observe which position is the edge in the test set. 
Despite Model 1 appearing as the optimal choice based on the F1-score, a substantial proportion of its predicted call edges rank 11th or lower among all candidates, with over 30\% ranking at 20 or beyond. This prevalence of potentially false call edges raises significant concerns about the practical usability of Model 1 in real-world scenarios. It is worth noting what exactly causes the discrepancy between the two sets of metrics. The graph on the left in Figure~\ref{fig:roc-ranking} is produced using a balanced set of true positive-true negative edges, while the one on the right considers all the potential edges for each call site.

As a result, we choose to evaluate our model based on the ranking of predicted call edges, offering a more comprehensive and realistic measure of their utility in practical applications. This approach acknowledges the realities of our concrete problem domain, providing a nuanced understanding beyond the limitations of conventional metrics. 
Formally, we establish the definition of rank as follows.
Let \( R_i \) represent the ranking of the \(i-th\) call site, and \( P_i \) represent the predicted probability (or score) assigned by the model to the i-th call edge being a valid edge. The ranking \( R_i \) is calculated as the relative position of the \(i-th\) call edge in the ordered list of predicted probabilities \( P_1, P_2, ..., P_n \), where \( n \) is the total number of possible call edges, for the call site of interest.
Mathematically, the ranking \( R_i \) can be expressed as:

\[ R_i = \text{Rank}(P_i) \]

Here, \(\text{Rank}(P_i)\) denotes the rank or position of the predicted probability \( P_i \) within the sorted list of all predicted probabilities. The smaller the rank, the higher the relative positive ranking of the call site. This definition captures the notion that the ranking of a predicted function definition is determined by how well its predicted probability compares to those of all other candidate functions, reflecting its relative position in the model's output.

\subsection{Implementation}
To implement~\toolname{}, we utilized various open-source tools, each serving a distinct role in our workflow. First, we use CodeQL~\cite{codeql} for static analysis, building custom queries to accurately extract call edges from the source code. We verified the validity of the query by contacting the CodeQL team and asking if the extracted call graph reflects the one used internally by the framework.
For dynamic analysis, we used Babel~\cite{babel}, a JavaScript compiler. We created a custom Babel pass along with a configuration file that enabled the automatic loading of this pass. This setup facilitated us to instrument the code and capture dynamic behaviors during runtime, which allowed us to extract call edges by logging the call stack at each function entry.
We implemented our graph neural network using the Deep Graph Library (DGL)~\cite{wang2019dgl} for graph construction and manipulation and PyTorch~\cite{imambi2021pytorch} for model development and training. In particular, our neural network leverages the GatedGCN architecture implemented by Dwivedi et al.~\cite{dwivedi2022benchmarking}. This implementation offers a versatile and ready-to-use framework suitable for various GNN architectures, providing a robust foundation for \toolname{}.
\section{Empirical Evaluation}
In this section, we first introduce our research questions and datasets, then present our empirical results and answer the research questions. Finally, we put our results into perspective with a discussion on how to use the proposed model.

\subsection{Research Questions}
Our evaluation aims to answer the following questions:

\textbf{RQ1:} \textit{Does \toolname{} effectively predict call edges in JavaScript graphs?} This inquiry assesses \toolname{}’s proficiency in ranking suitable callees on the basis of a static call graph generated using CodeQL. That is, we verify empirically if \toolname{} is powerful enough to approximate the computation performed by a state-of-the-art static call graph construction tool. The evaluation involves testing on a dataset comprising the 50 most popular large-scale npm libraries, focusing on \toolname{}'s capacity to rank true edges relative to other potential candidate edges.

\textbf{RQ2:} \textit{Can \toolname{} predict novel call edges?} This question delves into the system's capability to predict new edges that CodeQL, a static analysis tool, did not identify. Given the lack of automatic ground truth for these call sites, we resorted to dynamic analysis to obtain true edges that are beyond the reach of static analysis. This approach allows us to assess the ranking of these potential candidates and answer the question.


\textbf{RQ3:} \textit{Which components of \toolname{} contribute to its performance?} This investigation examines the utility of \toolname{}'s proposed code representation and features. Through an ablation study, we assess the contribution of individual components by selectively removing them and observing the resulting changes in \toolname{}'s performance.

\textbf{RQ4:} \textit{Can dynamic analysis enhance \toolname{}'s performance?} Considering \toolname{}'s reliance on underapproximate ground truth, this question explores the potential performance boost achievable by integrating additional call edges from dynamic analysis. The assessment involves incorporating dynamic edges during training and observing the impact on \toolname{}'s performance on test data.

\textbf{RQ5:} \textit{Can \toolname{} handle intricate scenarios such as higher-order functions, cross-file invocations, and indirect \texttt{apply()} or \texttt{call()} invocations?} This inquiry evaluates \toolname{}'s proficiency in addressing the complexities of the JavaScript language. The evaluation partitions the dataset into different subsets, and the rank of edges within each category is measured to assess \toolname{}'s effectiveness in handling diverse and complex cases.

\begin{table}
    \centering
    \begin{tabular}{ l|r r r r }
         \toprule
         \emph{Metric} & \emph{Maximum} &  \emph{Minimum} & \emph{Mean} & \emph{Median}\\
         \midrule
         Lines of Code (LOC)& 2.5M & 2K & 216K & 47K \\
         
         Number of Nodes & 705K  & 3K  & 84K & 44K\\
         
         Number of Edges & 1.1M & 5K  & 130K  & 67K\\
         
         Call Edges & 42K  & 271 & 2959 & 1034 \\
         \bottomrule
    \end{tabular}
    \caption{Summary statistics of the selected npm packages, including node and edge counts and lines of code.}
    \label{tab:stat}
    \vspace{-5mm}
\end{table}

\subsection{Dataset}
We carefully curated a dataset of 50 npm packages from the top 1000 most depended-upon packages~\cite{npmrank}, focusing on entries that each exhibited a significant level of functional interconnectivity. Our selection criteria emphasized packages with at least 250 non-built-in call edges identified through static analysis. This threshold was specifically chosen to ensure the model has enough data to learn from. The resulting dataset is substantial, encompassing approximately four million nodes and six million edges, with a notable 116K of these edges being non-built-in call edges that are key to understanding inter-function call dynamics. The description of our dataset is presented in Table~\ref{tab:stat}.

         
         
         

\subsection{Effectiveness of \toolname{}}
\label{rq1}

\begin{figure*}
\begin{center}
  \includegraphics[width=0.85\linewidth, keepaspectratio]{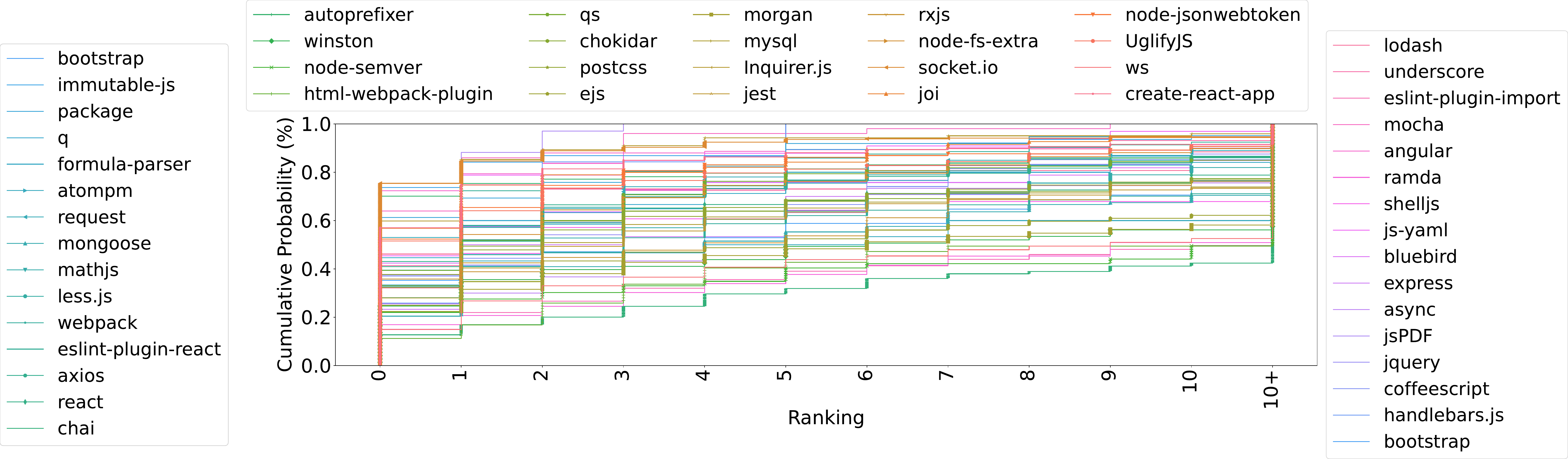}
\end{center}
\vspace{-3mm}
\caption{Distribution of Edge Prediction Ranking in \toolname{}. This graph illustrates the performance of \toolname{} on multiple npm libraries, in a purely static setup in which both training and testing data are produced by the baseline static analysis.}
\label{fig:rq1}
\vspace{-5mm}
\end{figure*}

To address RQ1, we independently train \toolname{} on each package in our dataset and aim to predict statically-known edges. For each call site to be resolved, the target callee function may be any function defined in the current project. The findings, illustrated in Figure~\ref{fig:rq1}, underscore the efficacy of \toolname{} in accurately predicting known edges through the utilization of statically generated call edges. To conduct our experiment, we employed a random data split, reserving 10\% for testing purposes. Notably, our test set exclusively comprises call edges derived from the output of the static analysis tool. The results demonstrate that \toolname{} can successfully approximate CodeQL's static analysis, successfully forecasting the true edge at rank 0 in 42\% of cases across the evaluated libraries. Additionally, in 72\% of instances, our model accurately predicts a new edge within the top five rank.

\begin{figure}
\begin{center}
  \includegraphics[width=\linewidth, keepaspectratio]{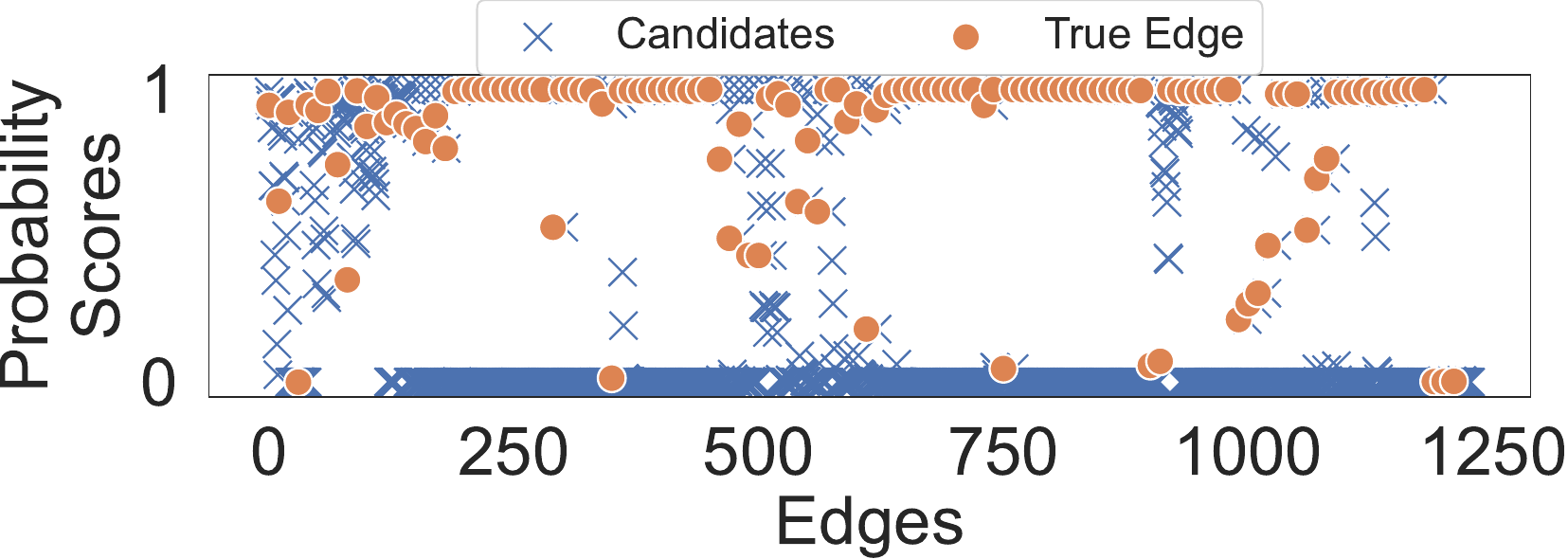}
\end{center}
\vspace{-3mm}
\caption{Confidence Score of Candidate Edges in \textit{express} - Each point on the x-axis is an individual call site, for which we plot the highest ten predictions. The true edges are depicted by orange circles, and other candidates by blue crosses.}
\label{fig:rq1_distribution}
\vspace{-5mm}
\end{figure}
These statistics are derived using a weighted average approach, where the influence of each library on the overall result is proportional to its size. This method ensures that larger or more complex libraries, which typically present more challenging prediction scenarios, have a corresponding impact on the average performance metric, providing a more accurate and representative assessment of the model's capabilities. While there is a large variance across the entries in the dataset, the results generally show that \toolname{} can handle a wide variety of real-world code and is able to resolve call sites under realistic scenarios, in which the callee function may be located anywhere in the project.

Figure~\ref{fig:rq1_distribution} presents the confidence score of 10 highest ranked predictions, for each call site in the \textit{express} library. These results clearly demonstrate that true edges, denoted by orange circles, are predominantly positioned at the top of the graph, indicating higher confidence in the prediction. In contrast, the blue crosses representing other candidates are mainly clustered at the bottom, with signifying lower probabilities. This distinct separation in the graph highlights \toolname{}'s effectiveness in distinguishing between true and false edges across all call sites.

\begin{tcolorbox}

\toolname{} successfully approximates the static call graph construction by
identifying statically-known edges at rank~0 in 42\% of cases. Furthermore, it can predict call edges within the top 5 rank in 72\% of instances. \toolname{} also tends to associate a high confidence score for true edges in our dataset.
\end{tcolorbox}

\subsection{\toolname{}'s Ability to Generalize}
\label{rq2}
\begin{figure}[H]
\begin{center}
  \includegraphics[width=\linewidth, keepaspectratio]{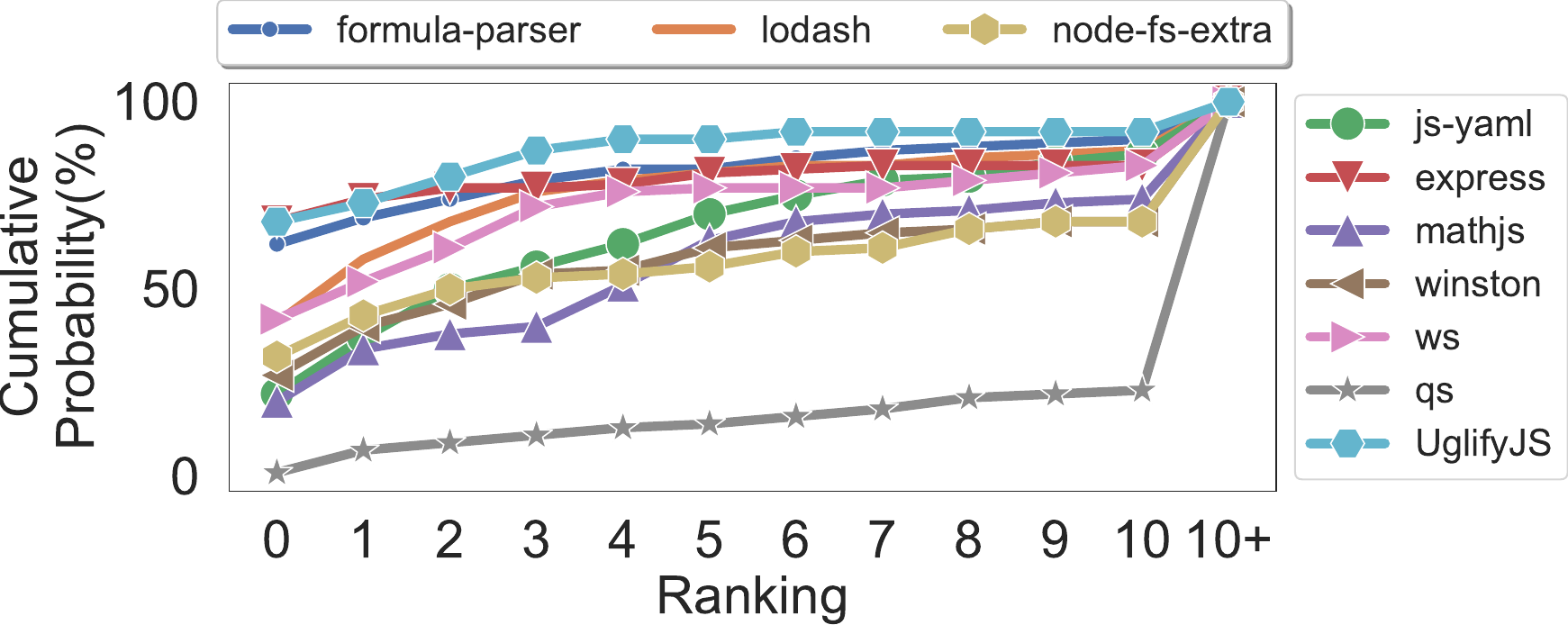}
\end{center}
\vspace{-3mm}
\caption{Distribution of \toolname{}'s Performance in Predicting Statically-Unknown Edges - We train the model with static ground truth, but evaluate it with dynamically-extracted ones.}
\label{fig:rq2}
\vspace{-3mm}
\end{figure}

To assess \toolname{}'s capability in predicting novel edges that elude static analysis, we conducted tests with unknown edges that the static analysis tool could not discover. To establish ground truth for performance evaluation, we employed a dynamic analysis approach using Babel-based instrumentation to extract dynamic function invocations. Utilizing available unit tests for each library, we focused on ten large npm libraries with substantial test coverage, recognizing that unit tests may not cover the entire code base, resulting in partial ground truth for dynamic call edges. 
We trained the model exclusively with static call edges, consistent with the setup in ~\ref{rq1}, and assessed the rank of each edge relative to other potential candidates. Figure~\ref{fig:rq2} visually represents the efficacy of \toolname{} in resolving call sites that CodeQL cannot handle.


In the case of the \texttt{express} and \texttt{UglifyJS} libraries, \toolname{} successfully predicts true edges in 68\% of cases at rank 0. However, in the most challenging scenario with the \texttt{qs} library, \toolname{} struggles to predict new edges at rank 0. We suspect this is the case because most of the call sites that are not solved by CodeQL in this project are from a JavaScript bundle file containing thousands of lines of code and intricate ways of invoking functions via an emulated require method. The model probably has few examples in the ground truth to aid it in handling such complex cases. 
Nonetheless, these results show that \toolname{} can predict call edges that are beyond the reach of existing static analyses, even when trained exclusively with ground truth produced by such analyses.

\begin{tcolorbox}
\toolname{} successfully predicts novel edges missed by static analysis, achieving a 68\% success rate for \texttt{express}. Furthermore, on average, it predicts almost 72\% of edges within the top~5 rank, demonstrating \toolname{}'s ability to analyze diverse code.
\end{tcolorbox}
\subsection{Ablation Study}
\label{sec:ablation}
\begin{figure}
\begin{center}
  \includegraphics[width=.9\linewidth, keepaspectratio]{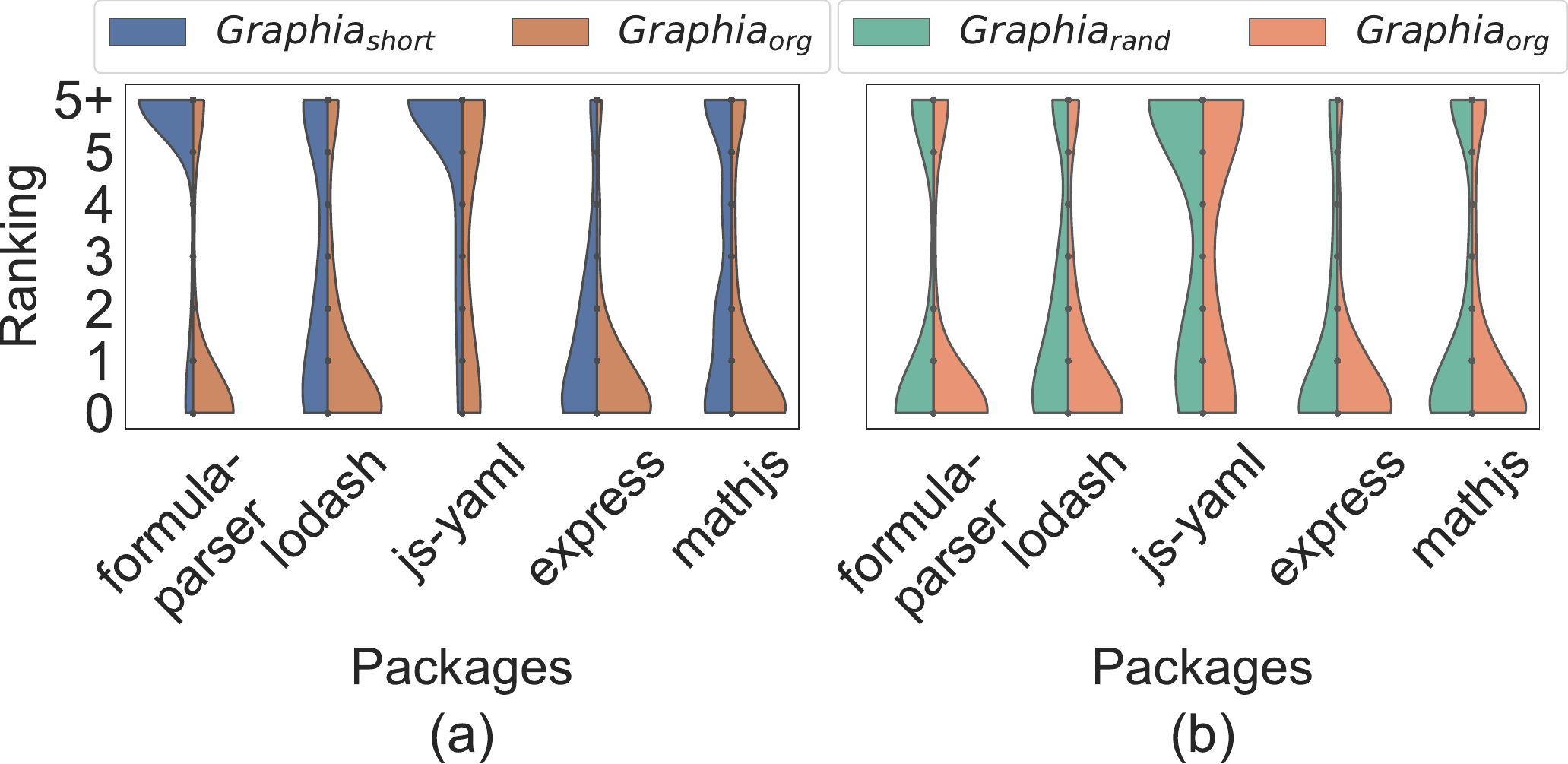}
\end{center}
\vspace{-3mm}
\caption{Performance Comparison: \toolname{} with Call Nodes (\(Graphia_{short}\)) vs. Original Design (\(Graphia_{org}\)) (a), and Performance Analysis: \toolname{} with Null Features (\(Graphia_{rand}\)) vs. Original Design (\(Graphia_{org}\)) (b).}
\label{fig:rq3}
\vspace{-5mm}
\end{figure}

To address this question, we conducted two distinct ablation studies focusing on the significance of the code representation and the proposed features.
In the first experiment, we assess the importance of our proposed code representation by removing the structure that binds certain code elements. 
Figure~\ref{fig:rq3}a presents the outcomes of these experiments, where \(Graphia_{short}\) denotes \toolname{} utilizing only \(FunctionExpression\) and \(CallExpression\) nodes along with their associated features, and \(Graphia_{org}\) represents \toolname{} in its original structure. As anticipated, the removal of the code structure has a substantial impact on the model's performance. The performance exhibited a 61\% drop for \texttt{formula-parser} at rank~0. Similarly, the model's performance drops by 31\%, 31\%, 18\%, and 23\% for \texttt{lodash, js-yaml, express, mathjs}, respectively.
In essence, if both syntactic and semantic structures are removed from \toolname{}, there is a noticeable decline in its performance, particularly in handling complex long-distance relations. The observed performance reduction suggests that, in the absence of a structured representation, graph neural networks struggle to infer intricate interprocedural relations, given the absence of a clear path for message passing. This underscores the critical role of both syntactic and semantic nodes and edges in ensuring \toolname{}'s effectiveness.


Next, we investigate the importance of the node features outlined in Table~\ref{table:features}. The results of this analysis are depicted in Figure~\ref{fig:rq3}b, where \(Graphia_{rand}\) and \(Graphia_{org}\) represent \toolname{} with null node features and the original structure, respectively. While the impact is less pronounced than the code structure ablation, the absence of node features still leads to notable differences, particularly at slightly higher ranks. Specifically, the performance experiences drop of 13\%, 10\%, and 8\% for \textit{formula-parser, js-yaml, express} respectively, when considering higher ranks. This underscores the importance of the proposed features, embedded in the nodes, suggesting that these features play a crucial role in \toolname{}'s effectiveness.

\begin{tcolorbox}
\toolname{}'s effectiveness significantly diminishes without syntactic and semantic structure. The model's output is further impacted when node features are absent, particularly at higher ranks, emphasizing their importance in the overall model's predictive capability.

\end{tcolorbox}
\subsection{Boosting Performance with Dynamic Edges}
\begin{figure}
\begin{center}
  \includegraphics[width=.95\linewidth, keepaspectratio]{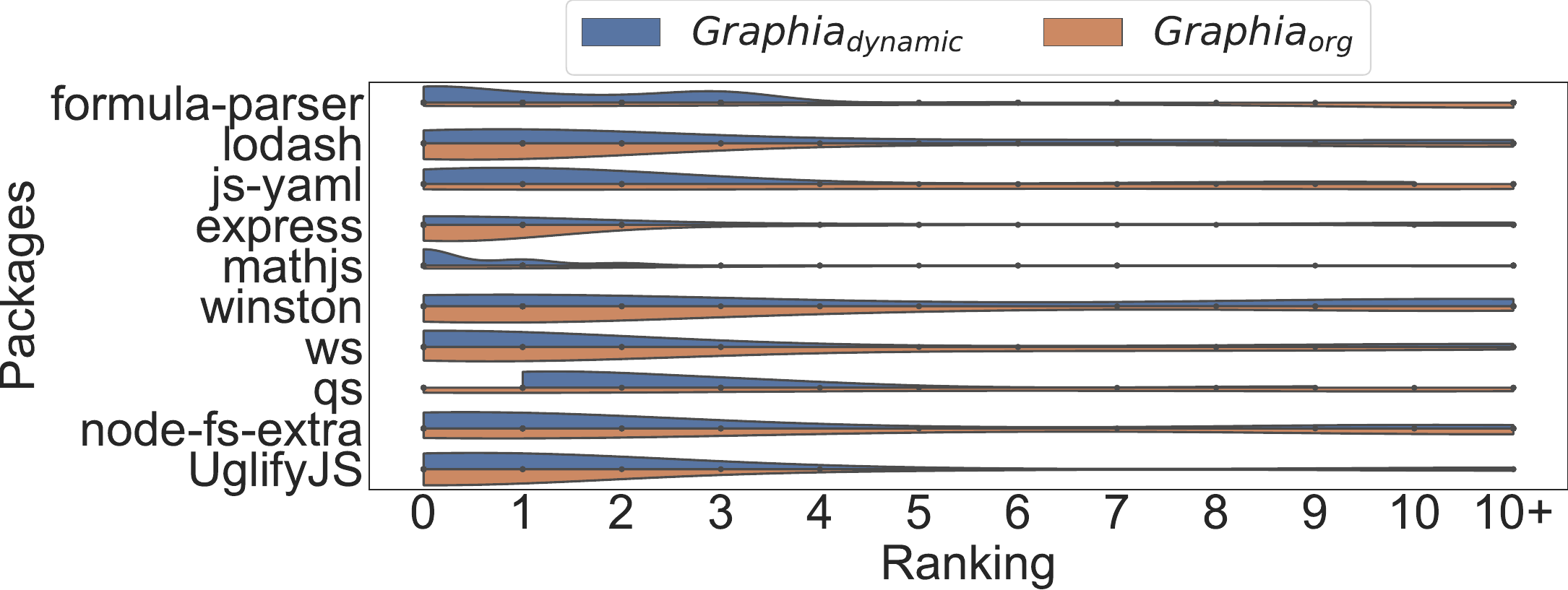}
\end{center}
\vspace{-3mm}
\caption{Impact of Integrating Dynamic Edges on \toolname{}'s Performance - The orange violin plots represent models trained solely with static ground truth, while the blue violins depict models additionally trained with dynamically-extracted edges.}
\label{fig:rq4}
\vspace{-5mm}
\end{figure}
In addressing the previous questions, our model was trained solely on static analysis call edges, yielding an under-approximated call graph. Here, we explore enhancing \toolname{}'s performance by integrating dynamic and static analysis. By merging dynamic and static call edges as per the approach detailed in \ref{rq2}, we aim to harness dynamic analysis's capability to capture intricate patterns beyond static analysis. Although attaining full coverage in dynamic analysis poses challenges, we hypothesize that our model can generalize to unexplored code segments.
In Figure~\ref{fig:rq4}, we compare statically trained models with hybrid ones incorporating dynamic edges for each package. Notably, for \texttt{formula-parser}, \texttt{js-yaml}, \texttt{express}, and \texttt{mathjs}, adding dynamic edges improved rank 0 predictions by 106\%, 50\%, 17\%, and 31\%, respectively. Conversely, \texttt{lodash} saw a nearly 40\% reduction at rank 0. However, dynamic edges had a more pronounced impact at higher ranks, with up to a 42\% improvement for \texttt{qs} at rank 5 and at least an 18\% increase for most libraries at the same rank. The variability in \toolname{}'s rank 0 performance stems from limitations in dynamic analysis. Although it offers insights into runtime behavior, it often generates a limited number of complex dynamic call edges due to coverage challenges. This data scarcity can lead to overfitting during model training, hampering generalization. Nevertheless, even partial dynamic analysis significantly complements incomplete static data, enhancing the overall model's performance.

\begin{figure}
\begin{center}
  \includegraphics[width=.9\linewidth, keepaspectratio]{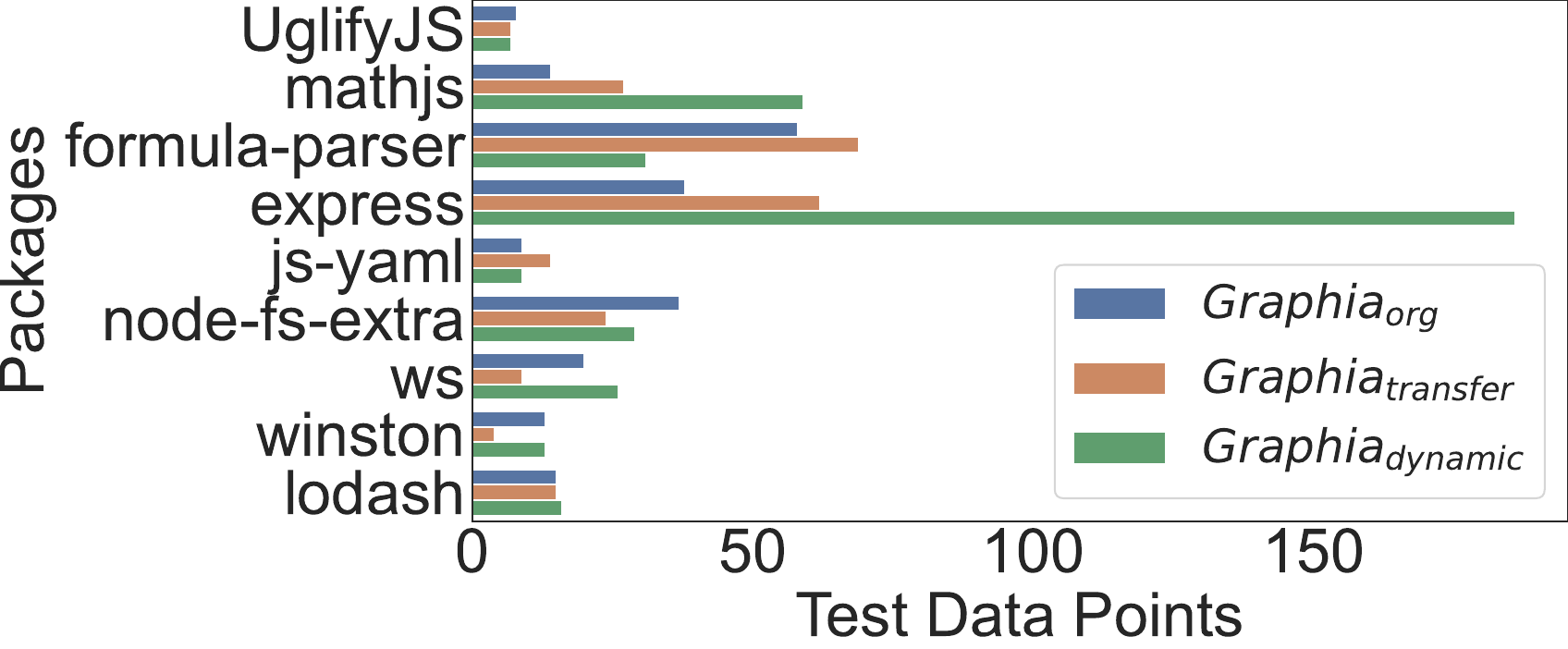}
\end{center}
\vspace{-3mm}
\caption{Impact of Transfer Learning on \toolname{}'s Performance. The blue bars represent models trained solely with static ground truth, while the green bars depict models trained with dynamically-extracted edges. Additionally, the orange bars signify models trained with data from four other projects.}
\label{fig:rq4_transfer}
\vspace{-5mm}
\end{figure}

To emphasize the impact of integrating dynamic edges, we conducted transfer learning experiments using a five-fold cross-validation approach. We trained models on combined static and dynamic edge graphs from four projects, relying only on static edges from the fifth. Figure~\ref{fig:rq4_transfer} shows that adding dynamic edges significantly boosts \texttt{mathjs}, \texttt{formula-parser}, \texttt{express}, and \texttt{js-yaml} performance at rank 0 by 92\%, 18\%, 63\%, and 55\% respectively. This shows the potential for performing dynamic analysis for only a handful of tested projects and reusing this knowledge to augment call graph construction for other projects.

\begin{tcolorbox}
\toolname{}'s integration of dynamic call edges showed mostly positive effects. For example, it notably enhanced rank 0 predictions for half of the considered programs by up to 106\%, with improvements reaching as high as 54\% within rank 5.

\end{tcolorbox}
\subsection{Dataset Partitioning: Predicting Hard Edges}

\begin{figure}
\begin{center}
  \includegraphics[width=\linewidth, keepaspectratio]{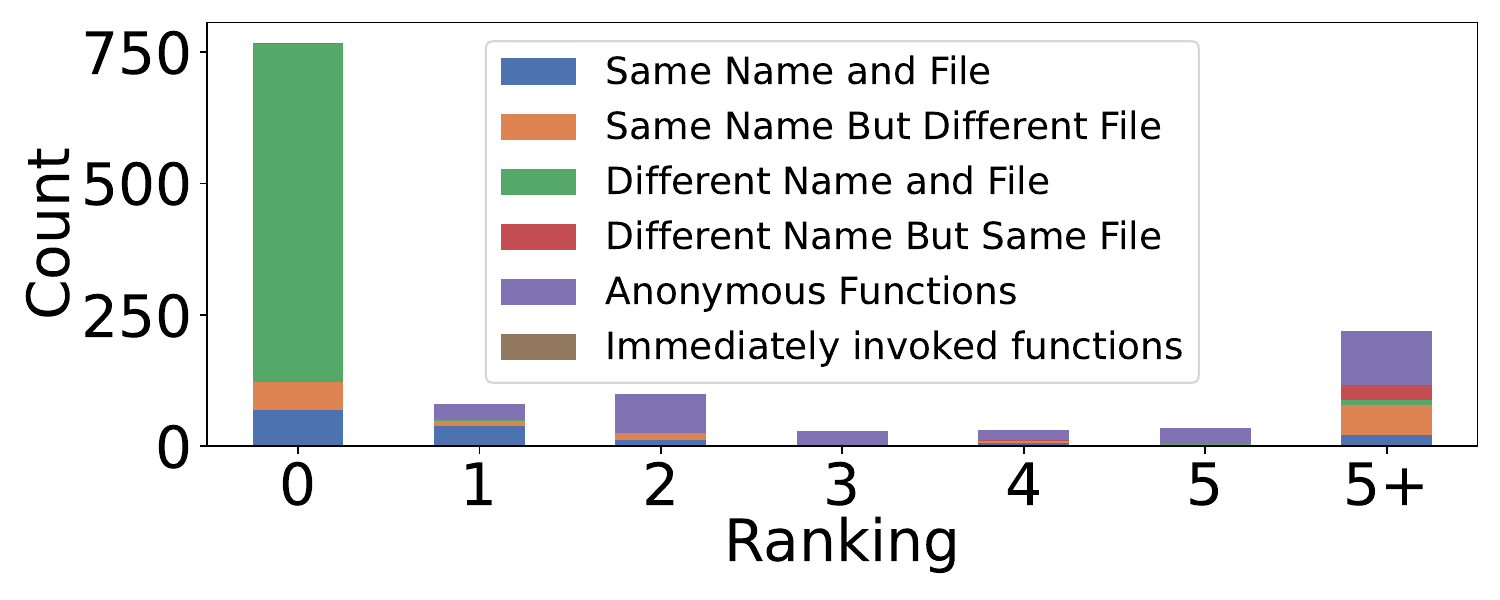}
\end{center}
\vspace{-3mm}
\caption{ Efficacy of \toolname{} in Predicting Call Edges Across Various Edge Types in JavaScript - Highlighting a 98\% success rate at rank 0 for functions with different names across files and a 63\% success rate in identifying anonymous function calls within rank 5. This figure depicts call sites from the \texttt{express} package.}
\label{fig:rq5}
\vspace{-5mm}
\end{figure}

Treating all call edges uniformly does not give any insight into why some edges are not resolved by the static analysis in the first place and into understanding the failure modes of the model. As noted earlier, there are many difficult-to-handle JavaScript language constructs that lead to a failure in resolving call sites. In this research question, we aim to partition the dataset based on the call site type and study the model's performance on each partition. 
We consider six distinct categories, facilitating a comprehensive evaluation of model performance. The assessment, as depicted in Figure~\ref{fig:rq5}, focuses on the specific case of \textit{express} package, revealing \toolname{}'s ability to infer relations even across multiple files successfully. 


Notably, in test scenarios involving functions with distinct names located in separate files, \toolname{} impressively predicts 98\% of new edges at rank 0. This is vital in cases where functions are defined as function expressions assigned to variables and later invoked through aliases, a common practice in JavaScript. Such non-local, distant relationships are challenging for static analysis tools to detect. Additionally, \toolname{} shows remarkable performance with anonymous functions, correctly predicting 63\% of new edges within rank 5. This indicates its strength in handling anonymous function calls, where static analysis tools often struggle. These results highlight \toolname{}'s versatility and effectiveness in dealing with the complexities of JavaScript code, particularly in scenarios involving higher-order or anonymous functions that pose significant challenges for traditional static analysis methods. The model's successful handling of various difficult edge types vouches for its applicability to real-world scenarios, where JavaScript code bases are typically rich in diverse functional constructs.

 
\begin{tcolorbox}
The considered case study shows that \toolname{} can handle hard features of the JavaScript language. For cases where the call site and the callee are in different files, \toolname{} places the correct function definition at rank 0 in 98\% of the cases. For calls to anonymous functions, \toolname{} places the correct function definition in the top 5 ranks in 63\% of the cases.

\end{tcolorbox}
\subsection{Threats to Validity}
The primary threat lies in using CodeQL to establish ground truth for training Graph Neural Network (GNN) models. Although effective, CodeQL can introduce false negatives into the derived ground truth, and, more rarely false positives. This limitation may lead to a likely underestimation of the actual prevalence of call edges, potentially compromising the effectiveness of GNN models and biasing the training. Consequently, their ability to generalize accurately to real-world scenarios may be hindered. While achieving a sound and complete static analysis is a challenging task, leveraging multiple static analysis tools alongside CodeQL could produce more reliable training data understanding of the data, thereby mitigating the impact of false negatives and false positives. 

The second threat to validity concerns the generalizability of the results to programs beyond our dataset. We attempted to mitigate this threat by considering a large set of popular, real-world libraries of various sizes.
\subsection{Discussion}

The presented results show that \toolname{} can often predict edges that are beyond the reach of existing static analyses. However, fully integrating this approach into a static analysis pipeline is far from trivial. The lack of GNN's explainability 
might lead to hard-to-debug false positives. Hence, we advocate for a human-in-the-loop approach where analysts vet the ranked list of candidate function definitions for unresolved call sites. To this end, future work should devise a way to identify the most valuable unresolved call sites that could lead to new findings in downstream analyses, if resolved. 

Another way in which the confidence in the prediction can be increased is to combine \toolname{} with additional signals, potentially coming from lightweight static analyses. For example, by employing a simple scoping analysis, \toolname{} might be able to reduce the list of potential target functions for a given call site, i.e., rule out functions that are not in scope.


\section{Related Work}
This section extends the discussion of prior work sketched in the introduction to discuss additional uses of machine learning in program analysis.

\paragraph{Applications of Machine Learning to Static Analysis}
Many studies have employed machine learning to enhance static analysis, particularly in addressing false positives~\cite{utture2022striking, le2022autopruner, tripp2014aletheia, flynn2018prioritizing, mir2024effectiveness}. The common approach involves using a static analysis tool to collect error reports, manually labeling them, selecting a feature set for the data, and training a classifier to identify false positives. While the overall methodology is consistent, the choices in bug-reporting tools, datasets, and feature sets vary, demonstrating adaptability to different scenarios in static bug analysis. For example, Utter et al.~\cite{utture2022striking} employed hand-picked semantic features and a Random Forest model to reduce false positives in call graphs for Java programs. Thanh et al. ~\cite{le2022autopruner} utilized a transformer-based approach to capture code embeddings and semantic features for false positive reduction. Tripp et al.~\cite{tripp2014aletheia} introduced ALETHEIA, a user-centric approach using statistical learning to refine static security analysis output based on user feedback. Flynn et al. ~\cite{flynn2018prioritizing} proposed an alert triaging tool that combines multiple classifiers, historical audit data, and hand-picked features.
In contrast, our work distinguishes itself by aiming to construct the call graph from scratch rather than pruning an existing one. Furthermore, our approach considers the entire code structure, a factor not addressed in previous studies.

\paragraph{Code Embedding Using Deep Learning}
Deep learning techniques have been widely applied to generate code embeddings, which serve as foundational representations for various downstream tasks. These tasks span a range of applications, including type inference~\cite{allamanis2020typilus, pradel2020typewriter, hellendoorn2018deep,peng2022static, wang2021codet5}, code completion~\cite{Devanbu2012OnTN, Raychev2014CodeCW, svyatkovskiy2019pythia}, code clone detection~\cite{saini2018oreo, liu2023learning}, program repair~\cite{chen2019sequencer, mashhadi2021applying}, fixing bugs~\cite{dinella2020hoppity}, optimization~\cite{cummins2020programl}, fault localization~\cite{nguyen2022ffl, lou2021boosting}, among other analyses on source code~\cite{brown2020language, leclair2019neural,nguyen2022vulcurator}. Researchers have explored the use of deep learning in code embedding for diverse purposes. Notable examples include the work of Allamanis et al. ~\cite{allamanis2020typilus} and Pradel et al. ~\cite{pradel2020typewriter}, who leveraged deep learning for type inference in projects like Typilus and Typewriter. Hellendoorn et al. ~\cite{hellendoorn2018deep} and Peng et al. ~\cite{peng2022static} also utilized deep learning for type inference and static analysis. In the realm of code completion, Devanbu et al. ~\cite{Devanbu2012OnTN}, Raychev et al. ~\cite{Raychev2014CodeCW}, and Svyatkovskiy et al. ~\cite{svyatkovskiy2019pythia} employed deep learning techniques to enhance code suggestion and completion mechanisms. Code clone detection has also benefited from deep learning, with Saini et al. ~\cite{saini2018oreo} introducing the Oreo model for effective clone detection. Program repair, a critical aspect of software maintenance, has seen advancements through deep learning. Chen et al. ~\cite{chen2019sequencer} and Mashhadi et al. ~\cite{mashhadi2021applying} developed Sequencer and applied deep learning for program repair tasks. Fault localization, crucial for debugging, has been addressed by Nguyen et al. ~\cite{nguyen2022ffl} and Lou et al. ~\cite{lou2021boosting} using deep learning approaches. Analyzing source code from various perspectives has also been explored. Brown et al. ~\cite{brown2020language}, LeClair et al. ~\cite{leclair2019neural}, and Nguyen et al. ~\cite{nguyen2022vulcurator} employed deep learning for language understanding, neural code search, and vulnerability curation, respectively.


Feng et al.~\cite{feng2020codebert} introduced CodeBERT, a pre-trained model to enhance code-text tasks, particularly in code search. Karampatsis et al.~\cite{karampatsis2020scelmo} incorporated contextual embeddings into a JavaScript corpus to improve program repair by utilizing context-specific information. Guo et al.~\cite{guo2021graphcodebert} introduced Graphcodebert, which leverages data flow during pre-training to enhance understanding of semantic relations in code. Alon et al. developed code2vec~\cite{Alon2018code2vecLD} and code2seq~\cite{alon2018code2seq}, tailored for method name prediction within code snippets. These models highlight the significant impact of deep learning in source code analysis, improving tasks like code search, program repair, and method name prediction.

As previously highlighted,  existing embedding approaches lack consideration for the entire code structure and often struggle to emulate interprocedural analysis. The limitations in existing methods underscore a gap in addressing the holistic understanding of code, particularly in scenarios involving complex structures and interprocedural dependencies. This gap emphasizes the need for more comprehensive approaches that can capture the intricate relationships and nuances within entire codebases, enabling more robust and accurate representations in the realm of source code analysis.

\paragraph{GNN for Program Analysis} Graph neural networks have gained significant prominence in handling various graph-structured data and have found widespread application in numerous general-purpose tasks. In the realm of code processing, researchers have leveraged GNNs to address various tasks, including code summarization~\cite{fernandes2018structured}, expression generation~\cite{brockschmidt2018generative}, code edition~\cite{yin2018learning}, and type inference~\cite{allamanis2020typilus}. Notably, Allamanis et al. [5] utilized GNNs with an augmented graph incorporating abstract syntax trees and data flow information to tackle downstream code processing tasks. However, similar to previous work, their approach focuses on function code snippets rather than capturing the entirety of the program, including multiple files.
\section{Conclusion}

In this paper, we show that link prediction with graph neural networks can be applied to entire JavaScript programs. We propose a code representation that increases the graph connectivity by connecting identifiers with the same name. We show that \toolname{}, our link prediction approach, can learn both from a purely static ground truth produced by existing static analysis tools and from dynamic edges generated by running the unit tests. The results show that \toolname{} can assist existing call graph approaches in generating hard edges involving dynamic property accesses or higher-order functions. Considering the promising results, we plan to experiment with alternative code representations, e.g., by integrating control flow information and with different downstream analyses.

\section{Data Availability}

We provide an extensive replication package containing all our scripts (for training the model, parsing the code, and running the analyses) and data (serialized graph representation for the considered packages and the set of all 163k call edges):

\begin{center}
    \href{https://figshare.com/s/a3af20774d388d5a9d86}{https://figshare.com/s/a3af20774d388d5a9d86}
\end{center}
\bibliographystyle{ieeetr}
\bibliography{references}

\end{document}